\documentclass[final,letterpaper]{appolbmod}

    \usepackage{latexsym,bm,amsmath,amssymb,amsfonts}
    \usepackage{epsfig,graphics,graphicx}
    \usepackage{makeidx}
    \usepackage{citesort}
    \usepackage{slashed}

    \newcommand{\dif}{{\rm d}}
    \newcommand{\abar}{\bar{\alpha}}
    \newcommand{\atpi}{\frac{\bar{\alpha}}{2\pi}}
    \newcommand{\del}{\partial}
    \newcommand{\dY}{\dif Y}
    \newcommand{\lan}{\langle}
    \newcommand{\ran}{\rangle}
    \newcommand{\rme}{{\rm e}}

    \newcommand{\Lam}{\Lambda_{{\rm QCD}}}
    
    \newcommand{\nn}{\nonumber\\}
    \newcommand{\ns}{&\!\!\!\!\!&}
    \newcommand{\mcal}{\mathcal}
    \newcommand{\order}[1]{\mcal{O}{(#1)}}

    \newcommand{\beq}{\begin{eqnarray}}
    \newcommand{\eeq}{\end{eqnarray}}

    \def\thenames{D.N.~Triantafyllopoulos}
    \def\thetitle{Running Coupling Effects in Small-$x$ QCD}
    \def\titleheading{\small \tt \hfill ECT$^*$-08-04}

\begin{document}

\title{RUNNING COUPLING EFFECTS IN SMALL-$x$ QCD\thanks{Based on lectures given at ``School on QCD, Low-$x$ Physics, Saturation and Diffraction'',
Copanello (Calabria), Italy, July 2007 [To appear in Acta Physica Polonica B].}}
\author{D.N.~Triantafyllopoulos
\address{ECT*, Strada delle Tabarelle 286, I-38050, Villazzano (TN), Italy\\{\tt dionysis@ect.it}}}
\maketitle

\begin{abstract}
We study effects of the running of the coupling in QCD at small Bjorken-$x$ and in particular the ones related to gluon saturation. After introducing the steps taken to the derivation of the next to leading order nonlinear evolution equation, we discuss the infrared sensitivity of the Pomeron intercept, the energy dependence of the saturation momentum and the appearance of geometrical scaling, and the dominance of the running coupling effects over the ones introduced by loops of Pomerons.
\end{abstract}

\section{Short introduction and outline}\label{SecIntro}

One of the main active fields of research in Quantum Chromodynamics is the study of its behavior in the high energy limit. In general, a scattering process is considered as a high energy one, when the square of the total energy $s$ of the colliding objects is much larger than the momentum transfer $Q^2$ between them. Then one hopes to approach the problem via analytical methods, since in this limit there is the
possibility of a large kinematical window $s \gg Q^2 \gg \Lam^2$
where one can apply weak coupling methods. In lepton-hadron deep inelastic scattering (DIS) the high energy limit is equivalent to the small Bjorken-$x$ limit since $x=Q^2/s$.

The BFKL (Balitsky, Fadin, Kuraev, Lipatov) equation \cite{BFKL}
is the starting point for any approach to the high energy limit of QCD. It resums the Feynman diagrams in perturbation theory which are enhanced by logarithms of the energy and when the equation is solved a total cross section growing as a power of the energy emerges. At least a posteriori this growth is not so surprising since at high energies the
wavefunction of a hadron contains a large number of partons, mostly gluons, due to the available phase space for virtual fluctuations and due to the triple-gluon coupling in QCD. For such a wavefunction description one mostly relies on the dipole picture \cite{AM94a} which gives the evolution with the energy of the dipole density (and of higher density moments) of a hadron.

However, even though QCD at high energy will be in general characterized by high densities and increasing cross sections, one needs to find a mechanism to tame the too steep increase as predicted by the BFKL equation. The gluon density at a given momentum should
saturate \cite{GLR} and never exceed a value of order $\mathcal{O}(1/\alpha)$ (modulo factors which are logarithmic in energy) and equivalently the scattering at a given impact parameter should not exceed unity. The BK (Balitsky, Kovchegov) equation \cite{Bal96,Kov99a} derived from QCD adds a nonlinear term to the BFKL equation leading naturally to the fulfillment of the saturation and unitarity constraints. The B-JIMWLK hierarchy \cite{JKLWab,Bal01,CGCab,Wei02} (Balitsky, Jalilian Marian, Iancu, McLerran, Weigert, Leonidov, Kovner) is a specific generalization of the BK equation, but it seems not to lead to different results \cite{RW04} and therefore we will not discuss it at all.

Saturation of parton densities might play a significant role at experiments in current and future colliders. Saturation models \cite{GBW99a,IIM04} and geometrical scaling \cite{GBKS01}, which is a consequence of BFKL dynamics in the presence of saturation \cite{IIM02,MT02,MP04a}, are consistent with the description of the small-$x$ DIS data at HERA and high-$p_{\bot}$ spectra in deuteron-gold collisions at RHIC are again explained by properties of saturation \cite{KJ06}. One expects the phenomenon to be more relevant at the LHC not only in proton-nucleus collisions but also in proton-proton ones, for example in the production of dijets separated by a large rapidity interval (Mueller-Navelet jets) \cite{MN87,MR06,IKT08}.

Even though the BK equation may give a correct qualitative description, it does not give the correct quantitative one and perhaps this is not so surprising, since it corresponds to a leading order approximation. And in fact there are two sources of large corrections, loops of Pomerons \cite{MS04,IMM05,IT05,KL05a} and next to leading order (NLO) contributions \cite{KWrun1,Balrun,BalChi08}. Both the BK and the B-JIMWLK equations do not properly describe the hadronic wavefunction in regimes where the density is low and fluctuations become important. Even though we are primarily not interested in this region of phase space, the evolution is nonlocal and thus is affected by these low-density high-momentum modes. Extra terms, which give rise to the formation of Pomeron loops, need to be added to the BK equation and these terms strongly modify the fixed coupling evolution even at the qualitative level. In practice, however, it just happens that when running coupling corrections are also taken into account, the Pomeron loop effects are delayed up to super-high energies \cite{RCPL}.

These lecture notes are based on just two one-hour presentations, and therefore we shall not be able to enter detailed calculations, nevertheless we will try to show the direction to all the steps that need to be taken. Furthermore, it is assumed that the reader is already familiar with the formalism, the concepts and the results at the level of the leading order approximation. Otherwise we refer the reader to either the original papers or to existing lectures and reviews \cite{AM01,IV03,Wei05,Tri05}. We divide the main body of the paper in four parts. In Sec.~\ref{SecNLO} we describe the efforts made towards the derivation of the NLO nonlinear equation and in particular the issues related to the argument of the running of the coupling. In Sec.~\ref{SecPom} we restrict ourselves to the linear equation and we show, through the Pomeron intercept evaluation in a simplified problem, how the evolution becomes sensitive to infrared physics. In Sec.~\ref{SecSat} we deal with the nonlinear equation and the energy dependence of the saturation momentum. We see how geometric scaling emerges and how physics becomes insensitive to the infrared behavior. Finally, in Sec.~\ref{SecLoop} we compare the effects of the running of the coupling to the ones introduced by loops of Pomerons, showing that the former dominate.

\section{Towards the NLO nonlinear equation}\label{SecNLO}

There are various versions of the BFKL equation depending on the ``observable'' or the quantity considered and the representation (momentum or coordinate space). For example one can consider the amplitude in quark-quark scattering (in momentum space), or in dipole-dipole scattering (in coordinate space), or the gluon density in a hadronic wavefunction, or the dipole density in a heavy onium wavefunction. When we take into account the nonlinear effects, there is presumably a unique route to follow, since the corresponding evolution equations acquire a relatively simple form only when we consider the problem of the scattering of a small in size color dipole off a generic hadron. For this particular quantity it is also easy to give a (not rigorous) derivation of the nonlinear equation, which also serves as a benchmark for the derivation of the one at next to leading order.

\begin{figure}[t]
\begin{center}
\includegraphics[width=0.99\textwidth]{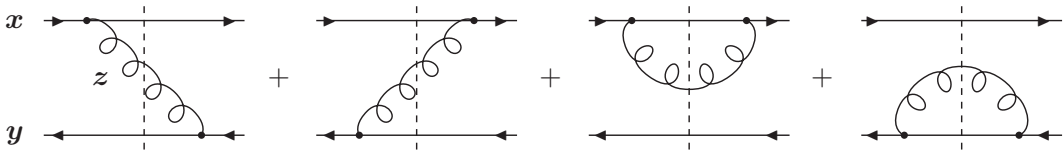}
\caption{\sl Soft gluon emission from a color dipole.}
\label{FigLO}
\end{center}
\vspace*{-0.25cm}
\end{figure}

Let us assume that we are in a frame where most of the energy is carried by the hadron, so that the color dipole ($\bm{x},\bm{y}$), with $\bm{x}$ and $\bm{y}$ the corresponding coordinates of the quark and the antiquark,  is ``bare''. That is, its wavefunction does not contain any higher order components. When the total energy, or equivalently the rapidity difference between the two colliding objects increases, we prefer to give the extra amount of energy to the dipole. Then its wavefunction evolves in a way that we can follow and in fact calculate. If $\dY=\dif k_{+}/k_{+}$ is the rapidity increment, then, to lowest order in $\alpha \dY$, either the quark or the antiquark emits a soft a gluon,  with $k_{+}$ its longitudinal momentum. By taking into account the four diagrams in Fig.~\ref{FigLO}, we can calculate the differential probability for the emission of the soft gluon in the interval $\dif Y\dif^2\bm{z}$, where $\bm{z}$ is the gluon position. We find \cite{AM94a}
   \beq\label{SplitProb}
    \dif P =
        \atpi\,
        \frac{(\bm{x}-\bm{y})^2}
        {(\bm{x}-\bm{z})^2(\bm{z}-\bm{y})^2}\,
        \dif^2 \bm{z}\, \dY
        \equiv
        \atpi\,
        \mathcal{M}_{\bm{x}\bm{y}\bm{z}}\,
        \dif^2 \bm{z}\, \dY,
    \eeq
with $\abar = \alpha N_c/ \pi$ and where $N_c$ is the number of colors. In the multicolor limit one can view the gluon as a quark-antiquark pair, and thus the state of the evolved ``parent'' dipole consists of two ``child'' dipoles ($\bm{x},\bm{z}$) and ($\bm{z},\bm{y}$). This two-dipole configuration scatters of the target and therefore the change $\dif S_{\bm{x}\bm{y}}$ in the $S$-matrix for the dipole-hadron scattering will be equal to $\int_{\bm{z}}\dif P S_{\bm{x}\bm{z}} S_{\bm{z}\bm{y}}$. There is also a term which corresponds to diagrams where the emitted gluon is absorbed before the scattering takes place. This term which normalizes the dipole wavefunction is equal to $-\int_{\bm{z}}\dif P S_{\bm{x}\bm{y}}$ and we arrive at the first Balitsky equation \cite{Bal96}
   \beq\label{Bal1}
    \frac{\del S_{\bm{x}\bm{y}} }{\del Y}=
    \atpi\,
    \int\limits_{\bm{z}}
    \mcal{M}_{\bm{x}\bm{y}\bm{z}}
    \left(S_{\bm{x}\bm{z}}S_{\bm{z}\bm{y}}
    - S_{\bm{x}\bm{y}}\right).
   \eeq
In principle, one should perform an average of all terms in the above equation over the target hadron wavefunction. With the mean field approximation $\lan S_{\bm{x}\bm{z}}S_{\bm{z}\bm{y}}\ran= \lan S_{\bm{x}\bm{z}} \ran \lan S_{\bm{z}\bm{y}}\ran$ we obtain a closed equation, the Kovchegov equation \cite{Kov99a}. Thus, in accordance with the natural conventions we refer to \eqref{Bal1} as the BK equation. Throughout our discussion we shall assume this factorization to be valid, and in Sec.~\ref{SecLoop} we will try to examine whether or not this is a good approximation\footnote{Therefore for notational economy we do not write the average brackets.}. We immediately notice that $S=1$ is an unstable fixed point, since any small initial amplitude, defined as $T=1-S$, will start to grow, while $S=0$ is a fixed one which corresponds to the unitarity limit.

At this leading order approximation the value of the coupling $\abar$ is assumed to be a small, but unknown, fixed number. This undesirable freedom forces us to put a running coupling by hand, but we are immediately lead to ambiguities since the evolution kernel $\mcal{M}_{\bm{x}\bm{y}\bm{z}}$ is nonlocal in the transverse space, and therefore an infinite number of combinations of the parent and dipole sizes could appear as the scale of the argument of the coupling\footnote{Notice that this is in sharp contrast to the situation encountered in the DGLAP equations; in that case the object of interest is a parton distribution function $f(x,Q^2)$, with $Q^2$ a resolution scale. The evolution is local in the transverse space and thus $Q^2$ arises naturally as the scale in the argument of the running coupling.}. Thus there is no other way out than to proceed to the calculation of higher order corrections to the BK equation \cite{KWrun1,Balrun,BalChi08}. Nevertheless, we need to say in advance that in general we are interested in the physics around the saturation scale $Q_s$, defined as the borderline between the region where the BFKL equation applies and the region where nonlinear effects become important and we approach the unitarity limits. Then we expect the dominant behavior of the $S$-matrix (and other observables) to be determined by letting $\alpha \to \alpha(Q_s^2)$ in the BK equation, and where $Q_s$ itself needs to be evaluated from the same equation. If this was not the case, then one might worry about the consistency of the whole construction. But let us postpone this analysis until Sec.~\ref{SecSat} and return to our original task.

In order to see what kind of contributions we need to calculate, we expand the QCD running coupling around its value at some fixed scale $\mu$ as
   \beq\label{alphaseries}
     \alpha(Q^2) = \alpha_{\mu} - \alpha_{\mu}^2\, \beta \ln \frac{Q^2}{\mu^2} + \alpha_{\mu}^3\, \beta^2 \ln^2 \frac{Q^2}{\mu^2} - \cdots,
   \eeq
where $\beta = (11 N_c - 2 N_f)/12 \pi$ is the leading order QCD $\beta$-function with $N_f$ the number of flavors. In terms of the parameters of the theory the diagrams in Fig.~\ref{FigLO} are of order $\alpha_{\mu}$. Diagrams with a quark-loop will be of order $\alpha_{\mu}^2 N_f$, diagrams with two quark loops will be of order $\alpha_{\mu}^3 N_f^2$ and so on, and match with the order of the terms in Eq.~\eqref{alphaseries}. Then we sum all the $\alpha_{\mu} (\alpha_{\mu} N_f)^k$ terms for $k \geq 1$ and we let $-2 N_f \to 11 N_c - 2 N_f = 12 \pi \beta$ in order to account for the gluon loop conributions\footnote{The gluon loop contribution has already been calculated \cite{BalChi08}.}. Finally we can read off the scale in the argument of the coupling and in order to obtain that properly it should be clear from the above that we need to focus in logarithmic contributions, like the ones in Eq.~\eqref{alphaseries}, in the transverse space.

\begin{figure}[t]
\begin{center}
\includegraphics[width=0.4\textwidth]{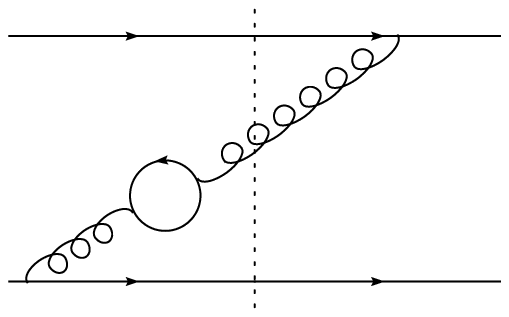}\hspace*{0.1\textwidth}
\includegraphics[width=0.4\textwidth]{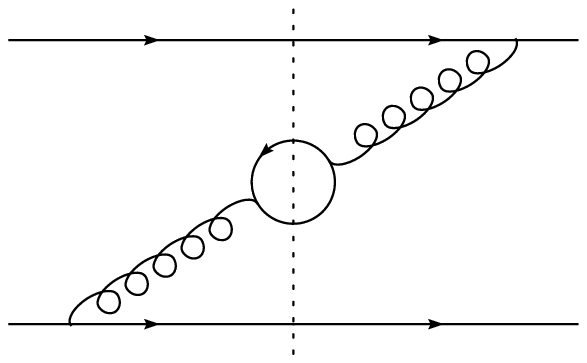}
\caption{\sl Left: Typical NLO diagram containing running coupling corrections. Right: Typical NLO diagram giving rise to a new state at the time of interaction.}
\label{FigNLO}
\end{center}
\vspace*{-0.25cm}
\end{figure}
There are two classes of diagrams of order $\alpha^2 N_f$ as shown in Fig.~\ref{FigNLO}. The first class, shown in the left panel of the figure, contains typical running coupling corrections; the soft gluon splits into a $q$-$\bar{q}$ pair which recombines before the time of interaction and thus the state at that time is the same as in the leading order case. Thus, from diagrams of this type we expect just a modification to the kernel of the leading order equation. In the second class of diagrams, shown in the right panel of the figure, the soft gluon splits into a $q$-$\bar{q}$ pair which does not recombine and therefore, when compared to the leading order, we have a new state at the time of interaction. Thus, the full NLO equation will have a more complicated structure which will involve a double two-dimensional integration over the transverse coordinates $\bm{z}_1$ and $\bm{z}_2$ of the quark and the antiquark of the pair. At a first glance it seems that these diagrams do not contribute to the running of the coupling, however this is not true as we will shortly see.

Let us first consider the ``simple'' diagrams of the first class. While we can integrate over the longitudinal momentum in the loop, the integration over the transverse momentum $\bm{k}^2$ is UV divergent. Using, for example, dimensional regularization and at the end letting $1/\epsilon \to \ln \mu^2$ \cite{KWrun1} we find the desired contribution to the running of the coupling.

Turning our attention to the more ``complicated'' diagrams of the second class let us see what happens when the pair shrinks to a point, which means that the loop transverse momentum becomes very large. Perhaps not unexpectedly, since it is hard to distinguish a zero-size pair from a gluon, we find that in this limit the diagram diverges and in fact this UV behavior also contributes to the running of the coupling. Therefore we subtract this divergent piece from the diagram in order to
obtain a UV finite result which corresponds to the new channel. Then we add it again to find the contribution to the running of the coupling.

The issue here is that there is not a unique way to do this separation of the infinities, since the ``point'' of subtraction can be chosen as a general combination of the quark and antiquark positions. Of course there is no issue regarding the NLO equation which is unique. Unfortunately it is not a closed equation; one also needs to write an evolution equation for the new state, which will involve another more complicated state, and so on. On the contrary, restricting ourselves to the running coupling contributions we obtain a closed equation which however is not unique. Since the task of deriving the NLO equation was pursued by two different groups two different schemes were used. Without any prejudice, but just because the equation is more compact we shall write here (part of) the result as given in \cite{Balrun}. In this ``B-scheme'' the NLO equation reads
   \beq\label{NLOBal}
    \frac{\del S_{\bm{x}\bm{y}} }{\del Y}=\ns
    \frac{\abar_{\mu}}{2\pi}\,
    \int\limits_{\bm{z}}
    \mcal{M}_{\bm{x}\bm{y}\bm{z}}
    \left[1 + \frac{\alpha_{\mu} N_f}{6 \pi} \,
    \ln \frac{\rme^{-5/3}}{(\bm{x}-\bm{y})^2\mu^2} + \cdots\right]
    \left(S_{\bm{x}\bm{z}}S_{\bm{z}\bm{y}}
    - S_{\bm{x}\bm{y}}\right)
    \nn
    \ns+ \abar_{\mu}\alpha_{\mu}N_f
    \int\limits_{\bm{z}_1 \bm{z}_2} [\textrm{new state}].
   \eeq
The second term in the square bracket of the first line corresponds to the running coupling contribution and one may already suspect that in this scheme the inverse of the parent dipole size sets the scale of the coupling (but not always as we shall shortly see). The second line corresponds to the formation and the interaction of the new state composed of the original quark and antiquark at $\bm{x}$ and $\bm{y}$ respectively, and the emitted quark and antiquark at $\bm{z}_1$ and $\bm{z}_2$. We will not deal with these new channel terms from now on.

The result of \cite{KWrun1} is slightly more complicated and the main difference when compared to the one given above in Eq.~\eqref{NLOBal} amounts to the replacement
  \beq\label{BtoKW}
    \ln \frac{1}{(\bm{x}-\bm{y})^2\mu^2} \to
    \ln \frac{R^2(\bm{r}_1,\bm{r}_2)}{\bm{r}_1^2 \bm{r}_2^2\mu^2},
  \eeq
where $R(\bm{r}_1,\bm{r}_2)$ is a known function of $\bm{r}_1= \bm{x} - \bm{z}$ and $\bm{r}_2= \bm{z} - \bm{y}$.

\begin{figure}[t]
\begin{center}
\includegraphics[width=0.4\textwidth]{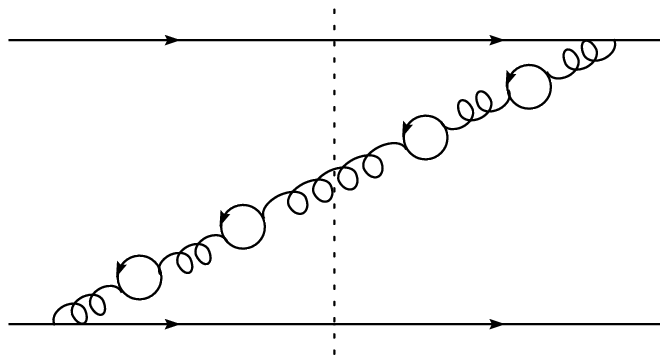}\hspace*{0.1\textwidth}
\includegraphics[width=0.4\textwidth]{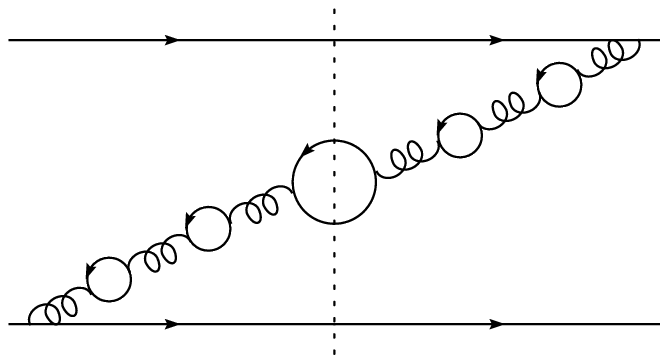}
\caption{\sl Typical bubble diagrams which need to be resummed in order to set the scale in the argument of the running coupling.}
\label{FigBubble}
\end{center}
\vspace*{-0.25cm}
\end{figure}

Finally we need to resum the bubble diagrams shown in Fig.~\ref{FigBubble} in order to obtain the structure given in Eq.~\eqref{alphaseries}. Notice that formally these diagrams correspond to N$^{n}$LO corrections with $n \geq 2$, but their resummation is equivalent to the setting of the scale in the coupling. At this point one may wonder about the number of resummations that we need to perform. The BFKL equation resumes $(\abar Y)^n$ enhanced terms, the nonlinear terms arise from the resummation of the high density effects in the target hadron wavefuntion, the bubble resummation is necessary to get the running coupling. Furthermore, we have to mention that one may need to perform a ``pole resummation'' since it is well known that the NLO BFKL kernel has a bad collinear behavior\footnote{And still, after all these resummations loops of Pomerons have not been taken into account.}. Thus one may be very enthusiastic about the high level of sophistication, but also may worry about the amount of control we have on the results after all these manipulations.

Taking into account the bubble diagrams we arrive at the nonlinear equation which, in the B-scheme, reads
  \beq\label{runBal}
    \frac{\del S_{\bm{x}\bm{y}} }{\del Y}=
    \frac{\abar(\bm{r}^2)}{2\pi}\,
    \int\limits_{\bm{z}}
    \left\{\mcal{M}_{\bm{x}\bm{y}\bm{z}}
    +\frac{1}{\bm{r}_1^2}
    \left[ \frac{\alpha(\bm{r}_1^2)}{\alpha(\bm{r}_2^2)} -1\right] + 1 \leftrightarrow 2
    \right\}
    \left(S_{\bm{x}\bm{z}}S_{\bm{z}\bm{y}}
    - S_{\bm{x}\bm{y}}\right),
   \eeq
with $\bm{r} = \bm{x} - \bm{y}$ the parent dipole size. In this B-scheme it is straightforward to see that the parent dipole size $\bm{r}$ sets the scale when $r_1=r_2$, while it is the smallest of the two child dipoles which sets the scale as $r_{<}$, when $r_< \ll r$ (so that $\bm{r}_> \simeq \bm{r}$). The latter is a very nice feature, since this is what we expect in the collinear limit. In the scheme of \cite{KWrun1}, and in view of Eq.~\eqref{BtoKW}, the r.h.s.~of the nonlinear equation is proportional to the triumvirate of running couplings
   \beq\label{trium}
     \frac{\abar(\bm{r}_1^2)\abar(\bm{r}_2^2)}{\abar(R^2)}.
   \eeq
Before closing this Section let us comment on possible problems that we may face because of the IR behavior of the coupling. So long as we are at fixed order $\alpha_{\mu}^2$, cf.~Eq.~\eqref{NLOBal}, large dipoles (with size bigger than $\Lam^{-1}$) need to be cut only in principle; they do not obstruct us to perform the integrations. When bubbles are resummed, cf.~Eq.~\eqref{runBal}, there is a non-integrable singularity. We need to introduce some type of an IR cutoff and, if we claim we have a sensible effective theory, we should be able to check cutoff-independence at the end. This will turn out to be true thanks to the dynamically generated saturation momentum $Q_s$. This scale becomes much larger than $\Lam$ at high energies, and therefore the r.h.s.~of the NLO equation is extremely small when we start to approach the pole at $\Lam$.

\section{Pomeron intercept and infrared sensitivity}\label{SecPom}

Restricting ourselves to the linear part of the evolution equation, that is to the BFKL equation with running coupling, we would like to find the behavior of the dipole-hadron scattering amplitude. More precisely, given a dipole of fixed size $\bm{r}$, we wish to find how fast the amplitude increases with rapidity. Here, and in the remaining Sections too, we shall neglect any dependence on the impact parameter of the process. In analogy to the fixed coupling problem \cite{BFKL} we expect an increase of the form $\exp(\omega_{\mathbb{P}} Y)$, with $\omega_{\mathbb{P}}$ to be determined. Just for reasons of simplicity in the presentation, we shall not try to deal with the running coupling BFKL equation, rather we will try to solve the equation
   \beq\label{diff}
      \frac{\del T}{\del Y} = \alpha(\rho)
      \left[1 + \left(\del_{\rho} + \frac{1}{2} \right)^2 \right] T
      \equiv \alpha(\rho)\, \mcal{K}\, T,
   \eeq
with $\alpha(\rho)$ the running coupling and where we have defined the logarithmic variable $\rho = \ln 1/\bm{r}^2 \Lam^2 $. We notice that the eigenfunctions $\exp(-\gamma \rho)$ of the operator $\mcal{K}$ are the same as the ones of the BFKL operator. The eigenvalue spectrum $\tilde{\chi}(\gamma) = 1+(\gamma-1/2)^2$ has roughly the same shape\footnote{In fact the spectrum of $\mcal{K}$ behaves better than the BFKL one, since the latter diverges for $\gamma \to 0,1$ violating energy conservation.} as the BFKL one, which we recall is given by $\chi(\gamma) = 2 \psi(1) - \psi(\gamma) - \psi(1 - \gamma)$ \cite{BFKL} with $\psi(\gamma) = \dif \ln \Gamma(\gamma)/\dif \gamma$; for real $\gamma$ both $\chi$ and $\tilde{\chi}$ are convex functions with a minimum at $\gamma=1/2$. We could choose some more general coefficients in Eq.~\eqref{diff} or even a more general form, however our final conclusion would not change.

With $\alpha=1/\rho$ (we let $\beta=1$ without any loss of generality) we can write the general solution to Eq.~\eqref{diff} in terms of the Airy function. We have
   \beq\label{diffsol}
    T(\rho,Y) = \sum_{\omega} c(\omega) \exp \left( \omega Y - \frac{\rho}{2}\right) \mathrm{Ai}\left( \frac{\omega\rho-1}{\omega^{2/3}}\right),
   \eeq
where $c(\omega)$ should be determined from the initial conditions. Now we need to enforce a boundary condition to cut the infrared contributions, e.g.~$T(\rho_0)=\mathrm{const.}$ with $\rho_0>0$, and, again for simplicity, we choose this constant to be zero. Then, for a given position of the boundary, $\omega$ can take only discrete values which are related to the zeros $-|\xi_n|$ of the Airy function. One needs to solve a transcendental equation to determine these allowed values of $\omega$, but for our purposes we can simply give them in the form of the series
   \beq\label{omega}
     \omega_n = \frac{1}{\rho_0} - \frac{|\xi_n|}{\rho_0^{5/3}} + \cdots = \alpha(\rho_0) - |\xi_n|\, \alpha^{5/3}(\rho_0) + \cdots.
   \eeq
Then our solution becomes
   \beq\label{diffsol2}
    T(\rho,Y) = \sum_{n=1}^{\infty} c(\omega_n) \exp \left( \omega_n Y - \frac{\rho}{2}\right) \mathrm{Ai}\left(-|\xi_n| +\omega_n^{1/3} (\rho-\rho_0)\right).
   \eeq
The rightmost zero of the Airy function at $-|\xi_1| = -2.33$ gives the largest value of $\omega$, the $n=1$ term dominates as $Y \to \infty$, and therefore $\omega_1$ as determined from Eq.~\eqref{omega} is the Pomeron intercept $\omega_{\mathbb{P}}$. The difference in QCD lies in the coefficients of the series expansion in Eq.~\eqref{omega} and the Pomeron intercept reads \cite{FR97}
   \beq\label{intercept}
    \omega_{\mathbb{P}} = 4 \ln2\, \abar(\rho_0) - |\xi_1|
    \left[\frac{\pi^2 \beta^2 \chi^2(\gamma_{\mathbb{P}}) \chi''(\gamma_{\mathbb{P}})}{2 N_c^2}\right]^{1/3} \abar^{5/3}(\rho_0) + \cdots,
   \eeq
with $\gamma_{\mathbb{P}} =1/2$, $\chi(\gamma_{\mathbb{P}}) = 4 \ln2$ and $\chi''(\gamma_{\mathbb{P}}) = 14 \zeta(3)$. It is a simple exercise to show that the $n=1$ term dominates up to values of $\rho$ such that $\rho - \rho_0 \lesssim \left[\alpha(\rho_0) Y \right]^{2/3}$, so that for large $Y$, $\rho$ can be in the perturbative region. On the one hand this is good, since the most one can achieve is to calculate the amplitude for dipole sizes much smaller than $1/\Lam$. On the other hand our solution cannot be trusted since it depends strongly on the cutoff; the Pomeron intercept is determined by the coupling which in turn is evaluated at $\rho_0$. Perhaps this should not come as a surprise, since Eq.~\eqref{diff} is similar to the Schr\"odinger equation with an attractive linear potential.

We could have imposed something ``milder'' than the absorptive boundary, like a coupling which freezes to a fixed value when reaching $\Lam$. Still, diffusion to the IR takes place and for any perturbative dipole $\rho \gg \rho_0$ the main contribution comes again from the region where the coupling is strongest, that is from momenta of order $\Lam$. Therefore BFKL evolution with running coupling is not self-consistent.

\section{The saturation momentum and geometric scaling}\label{SecSat}

\begin{figure}[t]
\begin{center}
\includegraphics[width=0.75\textwidth]{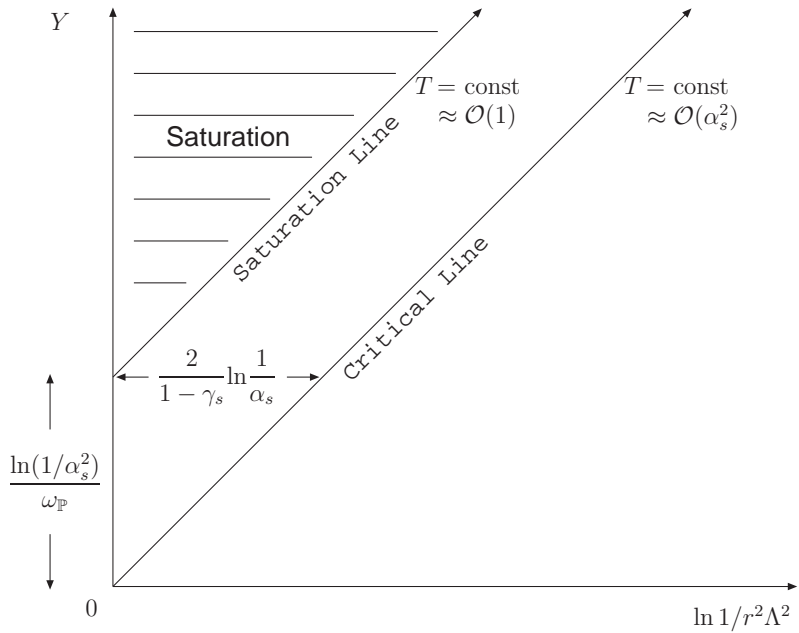}
\caption{\sl The saturation line in the logarithmic plane. With Pomeron loops included the evolution for determining the amplitude in the linear regime is restricted between the saturation and the critical line (see Sec.~\ref{SecLoop}).}
\label{FigPlane}
\end{center}
\vspace*{-0.25cm}
\end{figure}

Now we turn our attention to the problem of determining the saturation momentum $Q_s(Y)$, which can be defined as $T(\bm{r} = 1/Q_s(Y)) = \mathrm{const}.$ where the constant is of order $\order{1}$ but smaller than 1, as shown in Fig.~\ref{FigPlane}. In terms of the hadronic target wavefunction, it corresponds to the borderline between the low density momentum modes and the ones which are saturated.

In order to specify the energy dependence of $Q_s$ and the form of the amplitude for the scattering of dipoles with a size $r \lesssim 1/Q_s$ off the target hadron, it is enough to analyze the linear equations, but with appropriate boundary conditions which will play the role of the nonlinear effects \cite{MT02}. One needs to to be careful here since the boundary conditions are $Y$-dependent as may be suspected from Fig.~\ref{FigPlane}. Since the expectation is that the nonlinear terms will cut the diffusion to the IR, and more precisely to momenta (or inverse dipole sizes) smaller than $Q_s$, one needs to put an absorptive boundary just ``behind'' $Q_s$.

For specifying the leading behavior of $Q_s$ the detailed implementation of the boundary should not be crucial since the diffusion mechanism in the BFKL equation is an important but subdominant effect. Furthermore we expect that the physics for momenta around $Q_s$ line should be determined by $Q_s$ itself, otherwise our whole approach to the small-$x$ problem would not be very meaningful. Thus at the moment we will let $\alpha \to \alpha(Q_s)$ and under this replacement it is obvious that any scheme (as introduced in Sec.~\ref{SecNLO}) will lead lead to the same answer. Therefore we shall write our linear equation as
   \beq\label{linear}
     \frac{\del T}{\del Y} = \frac{1}{\beta \rho_s}\,
      \chi(1 + \del_{\rho}) T,
   \eeq
with the obvious notation $\rho_s = \ln Q_s^2/\Lam^2$ and the task is to find the line $\rho_s(Y)$ along which the amplitude $T$ is constant. There are four straightforward steps that we take: ({\tt i}) we change variable from $\rho$ to $z = \rho - \rho_s(Y)$, ({\tt ii}) we expand the function $\chi$ around the point $\gamma_s$ which is not known yet, ({\tt iii}) on the l.h.s.~of Eq.~\eqref{linear} we set the derivative of the amplitude w.r.t.~$Y$ equal to zero and, ({\tt iv}) we set the constant term and the coefficient of $\del_z$ on the r.h.s.~equal to zero. This last condition gives rise to two equations which determine both the anomalous dimension $\gamma_s$ and the saturation momentum $Q_s(Y)$. It is not hard to find that the leading $Y$-dependence of the saturation momentum is \cite{GLR,IIM02,MT02,MP04a}
   \beq\label{Qs}
     Q_s^2(Y) = \Lam^2 \exp \left[\sqrt{\frac{2 \chi(\gamma_s) N_c}{\pi \beta (1-\gamma_s)}\,(Y+Y_0)} \right]
   \eeq
with $Y_0$ an integration constant and where the anomalous dimension is given by
   \beq\label{chisol}
    \chi(\gamma_s) + (1-\gamma_s) \chi'(1-\gamma_s) = 0 \Rightarrow \gamma_s = 0.372,
   \eeq
which also leads to $\chi(\gamma_s)/(1-\gamma_s) = 4.88$. The form of the amplitude will be given shortly. We notice that, as a consequence of BFKL dynamics, $\gamma_s$ is a pure number. This number is smaller than $\gamma_{\mathbb{P}}=1/2$ which is the anomalous dimension corresponding to the line of fastest increase in the $(\rho,Y)$ plane (i.e.~the Pomeron intercept line), a fact which one could have anticipated by inspection of Fig.~\ref{FigPlane}.

A couple of comments should be made with respect to the behavior of the saturation momentum. The first is that it increases slower than in the fixed coupling analysis (recall that in the latter scenario the increase is exponential in $Y$ \cite{IIM02,MT02}). This is natural since the system evolves to higher rapidities along the saturation line; thus the momentum scale increases and the coupling decreases. We should perhaps mention here that Eq.~\eqref{Qs} is locally consistent with the fixed coupling result; indeed we have $\dif \ln (Q_s^2/\Lam^2)/\dY = [\chi(\gamma_s)/(1-\gamma_s)] \abar(Q_s^2)$. The second observation is that at very high energies $Q_s$ becomes the same for every hadron. For example in a large nucleus with atomic number $A$ one expects an enhancement of the saturation momentum by a factor of $A^{1/3}$. For not too high rapidities this is true, but this $A$-dependence which is hidden in the integration constant $Y_0$, becomes a subdominant effect for $Y \gg Y_0$.

Now we would like to go one step beyond and calculate the first preasymptotic correction to the saturation momentum given in Eq.~\eqref{Qs}. To this end, we need to expand any running coupling appearing in the nonlinear equation as
   \beq\label{alphaexpand}
   \alpha(\rho) = \frac{1}{\beta \rho_s}\,-\, \frac{z}{\beta \rho_s^2}.
   \eeq
Since different schemes correspond to different arguments of the running coupling, it is obvious that they will lead to different equations due to the second term of the expansion in Eq.~\eqref{alphaexpand}. Still, one can show that also the first correction to $Q_s$ is scheme-independent \cite{BP07}. Choosing a scheme, e.g.~the B-scheme where it is a bit easier to perform the calculation, we need to solve (approximately) a second order partial differential equation with $Y$-dependent boundary conditions. We find that the saturation momentum now reads \cite{MT02,MP04a}
   \beq\label{Qsrun}
     Q_s^2(Y) = \Lam^2 \exp \left[\sqrt{\frac{2 \chi(\gamma_s) N_c}{\pi \beta (1-\gamma_s)}\,(Y+Y_0)} - \frac{3 |\xi_1| A}{4} (Y+Y_0)^{1/6}\right],
   \eeq
while the scattering amplitude for the dipole-hadron scattering for $z>0$ (that is for dipoles such that $r \lesssim 1/Q_s$) reads \cite{MT02,MP04a}
   \beq\label{Trun}
    T(z,Y) = (Y+Y_0)^{1/6} \exp \left[ - (1-\gamma_s) z \right]
    \mathrm{Ai} \left(- |\xi_1| + \frac{z+c}{A (Y+Y_0)^{1/6}}\right),
   \eeq
where $A=\{[\chi''(\gamma_s)]^2 N_c/[2 \pi \beta (1-\gamma_s) \chi(\gamma_s)]\}^{1/6}$ with $\chi''(\gamma_s)=48.5$, $c$ is a constant of order $\order{1}$ and $Y_0$ is an integration constant which we will occasionally neglect from now on. Since the effect of the nonlinear term is to cut contribution coming from momenta smaller than $Q_s$, it is not surprising that the correction to $Q_s$ leads to an overall slower increase.

Now we notice in Eq.~\eqref{Trun} that within a distance $\sim Y^{1/6}$ (in logarithmic units) above the saturation line the amplitude becomes a function of a single variable $z=\ln 1/\bm{r}^2 Q_s^2$. More precisely, by letting $r \to 1/Q$ (just for illustration), the amplitude reads
   \beq\label{scaling}
     T = \left(\frac{Q_s^2}{Q^2}\right)^{1-\gamma_s}
     \left(\ln \frac{Q^2}{Q_s^2}\,+ c \right),
   \eeq
and we recognize the scaling form of the amplitude of the fixed coupling analysis, except that now $Q_s$ is different. It is important to realize that the scaling phenomenon persists even for momenta above $Q_s$, even though the diffusion radius, and therefore the region of validity of the scaling form, is now $\sim Y^{1/6}$ which is much smaller than the fixed coupling one ($\sim \sqrt{Y}$). The good thing about the smaller diffusion radius is that the evolution is less sensitive to ultraviolet contributions and therefore it is easier to perform a numerical study of the nonlinear equation. One should mention here that this scaling behavior is consistent with the interpretation of the small-$x$ data in electron-proton deep inelastic scattering \cite{GBKS01}, and, since there is no way to get geometrical scaling from the DGLAP equations, there is a hint that BFKL dynamics and saturation may have been observed.

\begin{figure}[t]
\begin{center}
\includegraphics[width=0.99\textwidth]{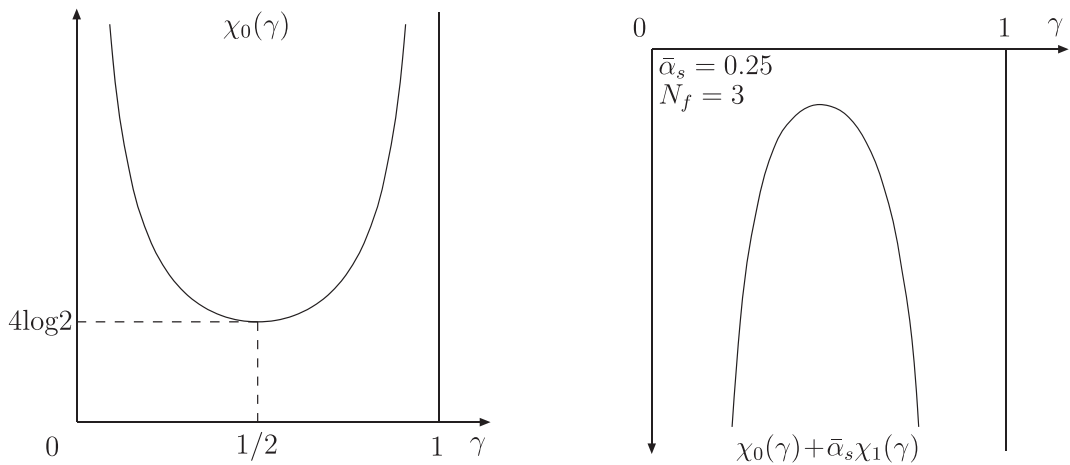}
\caption{\sl The characteristic function of the BFKL kernel as a function of the real part of $\gamma$ at leading order (left) and at next to leading order (right).}
\label{FigChi}
\end{center}
\vspace*{-0.25cm}
\end{figure}
Now one may ask the question what happens in the full NLO calculation. As we have done so far in this Section, let us consider NLO BFKL dynamics (for an introduction see \cite{Sal99}) together with the appropriate boundary conditions. The problem is that that the NLO kernel is unstable as one can see in Fig.~\ref{FigChi}; while the leading eigenvalue is positive, the NLO correction is negative and dominates\footnote{In fact this is true for real $\gamma$. The saddle point in the NLO case occurs at complex values of $\gamma$ leading to nonphysical oscillating cross sections.}. What happens is that the higher order corrections of the BFKL kernel do not behave properly in the collinear limit, and such collinear contributions should be resummed to all orders in the resummed perturbation theory \cite{Sal99}. Collinear physics is described by the DGLAP equations, and therefore one should ensure that the resummed BFKL kernel matches with DGLAP in the limits $\gamma \to 0,1$. If we neglect quarks, the matching condition reads
   \beq\label{match}
     \chi_{\rm{r}}(\gamma=0) = 1/\abar,
   \eeq
and similarly for $\gamma=1$. Eq.~\eqref{match} is simply equivalent to the energy conservation condition
   \beq\label{match2}
    \gamma(1) = 0 \qquad \mathrm{with} \qquad \gamma(\omega) = \int \dif z\,z^{\omega} P_{\rm gg}(\omega),
   \eeq
with $P_{\rm gg}(\omega)$ the gluon-gluon splitting function of the DGLAP equations. We shall not elaborate more into this, but simply say that the resummed kernel does not show any pathologies.

\begin{figure}[t]
\begin{center}
\includegraphics[width=0.75\textwidth]{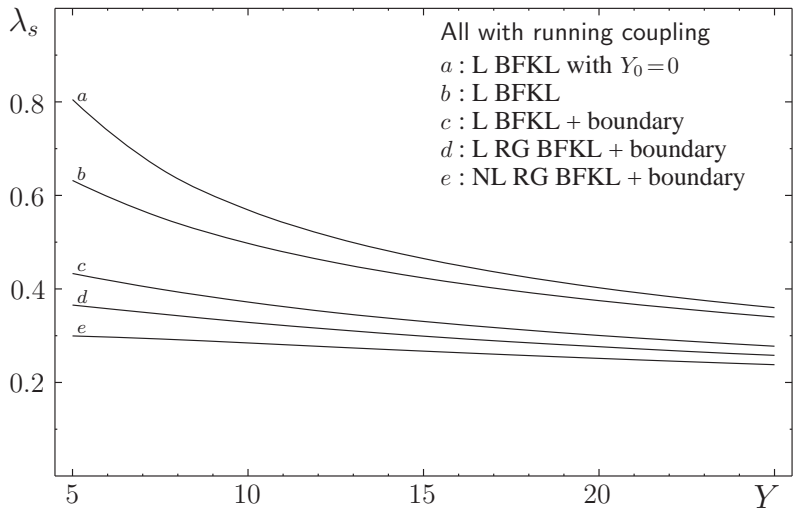}
\caption{\sl The logarithmic derivative of the saturation momentum $\lambda_s = \dif \ln (Q_s^2/\Lam^2)/\dY$ as a function of rapidity $Y$ for various kernels. Line-c corresponds to the running coupling result given in Eq.~\eqref{Qsrun} and line-e to the collinearly resummed NLO result.}
\label{FigLambda}
\end{center}
\vspace*{-0.25cm}
\end{figure}

Due to the complicated form of the resummed NLO kernel it is impossible to give an analytic expression for the saturation momentum. One observation is that at very high rapidity the full NLO result converges to the leading order result with running coupling \cite{Tri03}, since the coupling along $Q_s$ decreases. Defining the logarithmic derivative of the saturation momentum $\lambda_s = \dif \ln (Q_s^2/\Lam^2)/\dY$ we easily find from the analytic expression \eqref{Qsrun} that a typical value for $Y \simeq 10$ is $\lambda_s \simeq 0.4$. Taking into account the NLO corrections we indeed find a correction of order $\order{\alpha} \sim 30 \%$, as one would estimate, and therefore $\lambda_s \simeq 0.3$ as exhibited in Fig.~\ref{FigLambda} \cite{Tri03}. This is also what the fits based on QCD inspired saturation models give \cite{GBW99a,GBKS01,IIM04}.

\section{Running coupling versus Pomeron loop effects}\label{SecLoop}

Now we turn our attention to another type of corrections to the nonlinear equation. In order to motivate the introduction of these corrections, let us discuss some problems of the leading order evolution.

({\tt i}) The first problem is the extreme sensitivity to the ultraviolet. To understand the issue assume the coupling to be fixed and that we have evolved our system from zero rapidity up to rapidity $Y$ and we know the solution $T(r,Y)$. Now we try to reconstruct this solution by doing two (or more) global evolution steps; we evolve from zero to, say, $Y/2$ to obtain $T(r,Y/2)$ and then considering $T(r,Y/2)$ as an initial distribution we can evolve up to rapidity $Y$ to get $T(r,Y)$. We find that the solution obtained from this procedure agrees with the one obtained from the single global evolution step, only if we include (at least) the contribution from all dipoles such that $\ln1/r^2 Q_s^2 \lesssim \sqrt{D_{s}Y}$ in the initial condition at $Y/2$, with $D_s = 2 \abar \chi''(\gamma_s)$. There is no reason to cut the dipoles that lie outside the diffusion radius $\sqrt{D_{s}Y}$, but this algorithm reveals the width of phase space which is important for a self-consistent solution. The situation is quite embarrassing; with increasing $Y$, the phase space opens up to smaller and smaller dipoles, and the big numerical value of the coefficient $D_s$ makes the problem even worse. For instance, when one finds the saturation momentum to be a few GeV, at the same time one is sensitive to dipoles of inverse size a few orders of magnitudes above. This explains why in the numerical solutions, to both the BK and the JIMWLK equation, one had to go very far to the ultraviolet in order to obtain a reasonably accurate solution \cite{RW04}. In the running coupling case the situation is better, since the coupling decreases at higher momenta and thus the effects of these seemingly non-physical contributions are reduced. Indeed, as we saw in Sec.~\ref{SecSat}, the diffusion radius increases much slower, more precisely is proportional to $Y^{1/6}$. We shall come back to this later in this Section.

({\tt ii}) The second problem is the violation of unitarity. Say we want to calculate the amplitude close to, but above, the saturation line in the two ways we described in the previous paragraph. We have
    \beq\label{Tviolation}
    1 > c = T \sim \frac{1}{\alpha^2}\, T_{\rm a}\, T_{\rm b},
    \eeq
with $T_{\rm a}$ and $T_{\rm b}$ denoting the contributions of the two successive steps. It is clear that for $T_{\rm a} < \alpha^2$ the above equation imposes that the second step satisfy $T_{\rm b}>1$. Thus, all the paths going through the region to the right of the critical line
in Fig.~\ref{FigPlane} violate unitarity in the intermediate
steps \cite{MS04}. Returning to the problem we discussed in ({\tt i}), and noticing that the diffusion radius extends to the region where the amplitude can be much smaller than $\alpha^2$, we see that these contributions from the ultraviolet region must be indeed non-physical.

({\tt iii}) The successive emissions in the BFKL evolution lead to the formation of gluon cascades inside the hadron wavefunction. The nonlinear term in the BK equation corresponds to the merging of such cascades. Then one may wonder how could we have many of these (necessary for saturation) cascades. One possibility is that we have a large nucleus where there are many valence quarks and antiquarks and which serve as the sources for the generation of the gluon cascades. But of course this is just a particular initial condition and it does not offer the dynamical solution to the problem. One needs to find how QCD gives rise to the increase in the number of cascades and then one can start, for example, even from a single bare dipole and end up with a fully saturated wavefunction. We complete the theory by including the diagrams which were ``forgotten'' \cite{IT05} and which lead to the splitting of cascades. Such diagrams become important in the region $T \sim \alpha^2$ as needed in order to automatically solve the two problems presented in ({\tt i}) and ({\tt ii}). Since now we have both splitting and merging of cascades, we speak about loops of cascades, or loops of Pomerons.

Let us now estimate the effect of these loops of Pomerons in the saturation momentum $Q_s$ and the form of the amplitude around $Q_s$. For the moment we assume the coupling to be fixed. Since unphysical ultraviolet paths need to be cut, we will solve the BFKL equation with two absorptive boundaries, one in the infrared and one in the ultraviolet. Let $\Delta = (\ln 1/\alpha^2)/(1-\gamma_s)$ be the distance between the two boundaries in the logarithmic $(\rho,Y)$ plane, where, by definition, within $\Delta$ the amplitude drops from a value of order $\order{1}$ to a value $\order{\alpha^2}$. As usual, we make a change of variables from $\rho$ to $z= \rho - \rho_s(Y)$ and look for a $Y$-independent solution to the BFKL solution. It has to obey
   \beq\label{BFKL2b}
    [\chi(1 + \del_z) - \lambda_s\, \del_z]\, T(z) = 0,
   \eeq
where we recall $\lambda_s = \dif \rho_s/\dY$. The real combination of solutions which satisfies the boundary conditions is given by
  \beq\label{Tsol2b}
   T(z) \sim \exp[-(1-\gamma_{\rm r})z]\, \sin \frac{\pi z}{\Delta},
   \qquad \gamma_{\rm i} = \frac{\pi}{\Delta},
  \eeq
where $\gamma_{\rm r}$ and $\gamma_{\rm i}$ are the real and imaginary parts of $\gamma$. (We note that, in contrast to the single boundary problem, this solution does not correspond to any saddle point in the BFKL equation.) For a given value of $\alpha$, and therefore of $\Delta$ or $\gamma_{\rm i}$, the real part $\gamma_{\rm r}$ and the ``intercept'' $\lambda_s$ are uniquely fixed by
   \beq\label{lambda2b}
    \lambda_s = \frac{\chi(\gamma)}{1-\gamma}
    \qquad \mathrm{with} \qquad \mathrm{Im}(\lambda_s) = 0.
   \eeq
After we solve numerically the above transcendental equation, both the energy dependence of the saturation momentum and the amplitude are determined. In case the boundary separation is extremely large, or equivalently the coupling $\alpha$ is extremely small, Eq.~\eqref{lambda2b} leads to
   \beq\label{lambdaexpand}
    \frac{\lambda}{\abar} =
    \frac{\chi(\gamma_s)}{1-\gamma_s} -
    \frac{\pi^2 (1-\gamma_s) \chi''(\gamma_s)}{2 \ln^2 \alpha^2}.
   \eeq
Notice that the relative correction is proportional to $1/R_{\rm eff}^2$ with $R_{\rm eff} \sim \ln1/\alpha$ the effective transverse space for evolution, a feature which is true in general\footnote{For example one can easily check that this property holds in the running coupling expression \eqref{Qsrun}, with $R_{\rm eff} \sim Y^{1/6}$ being the diffusion radius.}.

The scaling behavior of the amplitude as given in Eq.~\eqref{Tsol2b} will not persist at very high values of rapidity because our system becomes stochastic. Pomeron loops modify the evolution in the region where the amplitude $T$ is of order $\order{\alpha^2}$ or equivalently the target hadron dipole density is of order $\order{1}$. In this low-density regime fluctuations become important and they lead to stochasticity. Thus different events lead to different profiles of the scattering amplitude as a function of $r^2$ and at a given fixed rapidity $Y$ \cite{IMM05}. These profiles are of the same form but shifted with respect to each other according to a probability density, which at a first approximation can be taken as a Gaussian with a width proportional to $\sqrt{Y}$. It is the averaging over all the events which leads to the violation of geometrical scaling. At this point it is useful to realize that the BK equation is deterministic and it corresponds to a Mean Field Approximation.

So now we are ready to reach our final goal. Both running coupling effects and Pomeron loop effects seem to be important and for practical purposes one cannot really rely on the leading order (fixed coupling) BK equation. Given the fact that there are no QCD evolution equations which include both effects, we are naturally forced to look if one of the two effects dominates. A first simple estimate seems to favor the Pomeron loops since, as we have just seen, they induce corrections which are of order $\order{1/\ln^2\alpha}$, while running coupling corrections are a part of NLO corrections\footnote{This is a bit naive estimate since bubble diagrams have been resummed to all orders.} which induce corrections of order $\order{\alpha}$. However, instead of these simple estimates, one would like to have a better control on such issues, for example by performing numerical solutions. This becomes a crucial issue, since the outcome will turn out to be not the expected one, that is, running coupling effects dominate the evolution.

\begin{figure}[t]
\begin{center}
\includegraphics[width=0.48\textwidth]{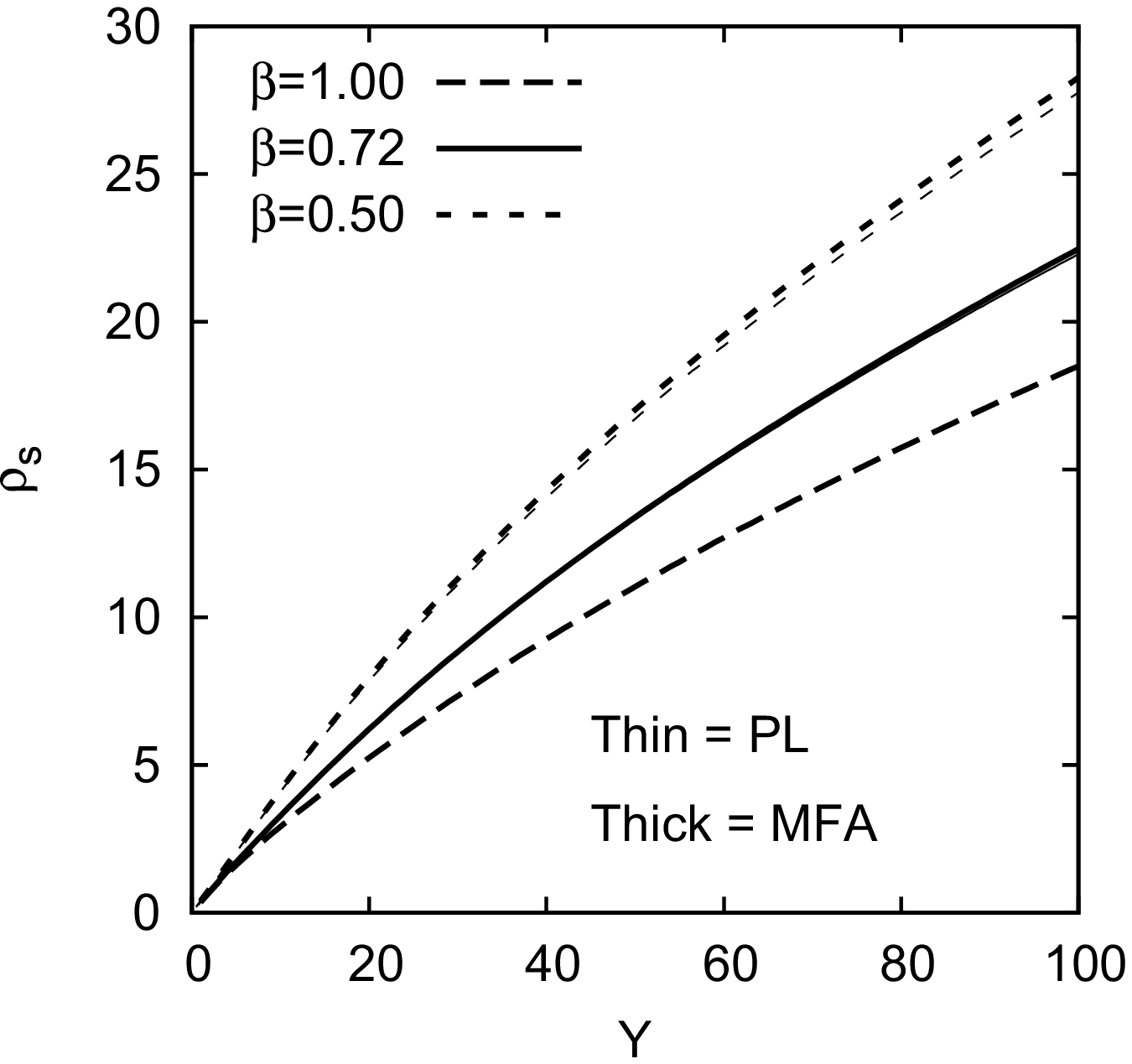}\hspace*{0.03\textwidth}
\includegraphics[width=0.48\textwidth]{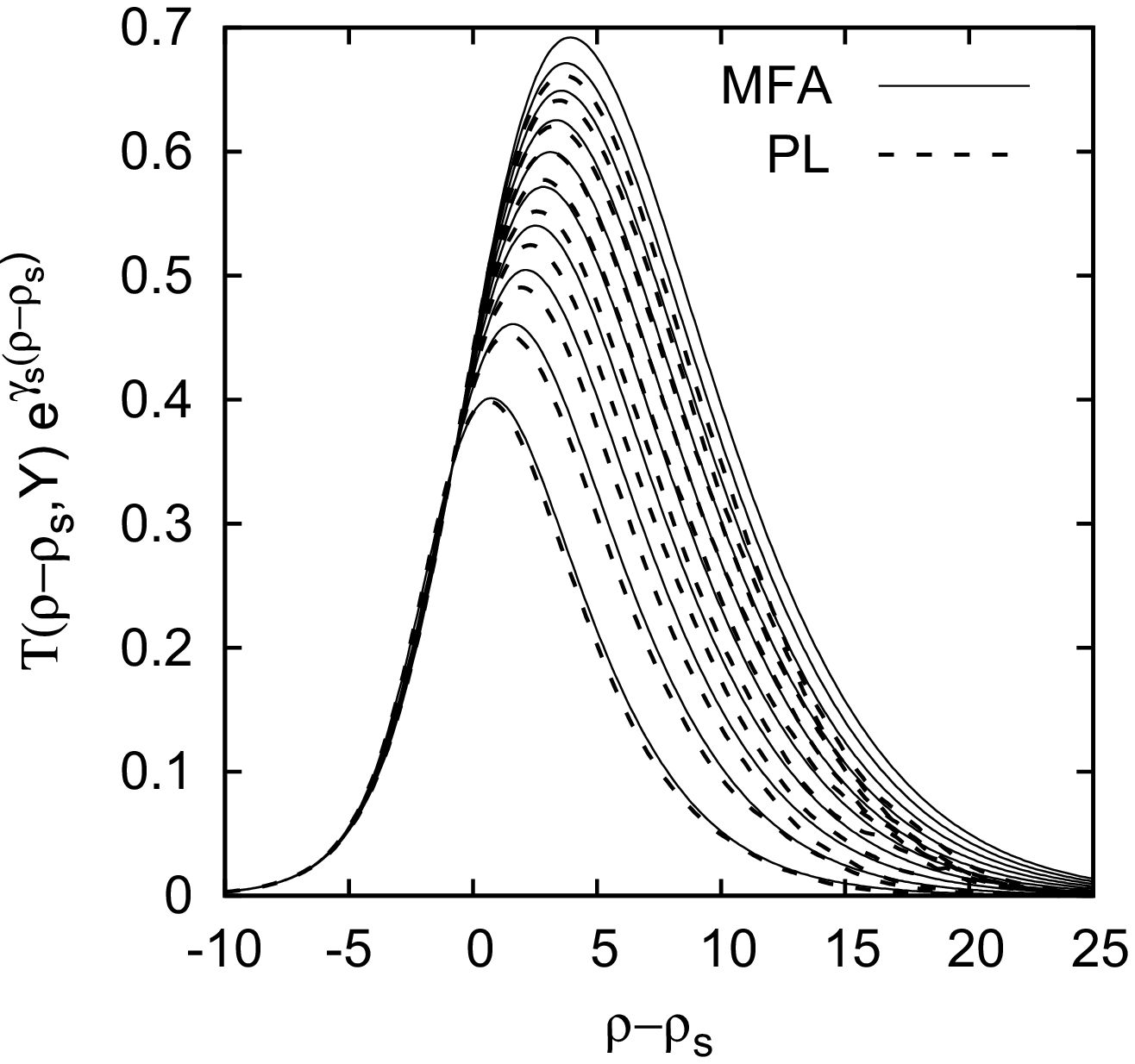}
\caption{\sl Left: The logarithm of the saturation momentum with Pomeron loops (PL) included and in the Mean Field Approximation (MFA) for various values of the $\beta$-function. One cannot distinguish between the two cases for the two largest values of $\beta$. Right: The reduced amplitude as a function of the logarithmic distance from the saturation line with Pomeron loops included and in the Mean Field Approximation.}
\label{FigSatT}
\end{center}
\vspace*{-0.25cm}
\end{figure}

Since we do not know the full effective theory, one way to proceed is to construct a model which contains both types of corrections, satisfies basic properties and principles of small-$x$ evolution (such as Lorentz invariance, emission of a single gluon under a step $\dY$ in rapidity, saturation of the emission rate at high gluon density,...) and is simple enough to be solved numerically. Such a model has been constructed \cite{RCPL}, and in the following we compare the results obtained from the numerical analysis of this model when ({\rm i}) both Pomeron loop and running coupling effects are included and ({\rm ii}) only running coupling effects are included. In the left panel in Fig.~\ref{FigSatT} we show the corresponding results for the logarithm of the saturation momentum $\ln Q_s^2/\Lam^2$ as a function of rapidity $Y$, and we see that there is no difference between the two cases up to super-high values of the rapidity. In the right panel in Fig.~\ref{FigSatT} we show the corresponding results for the ``reduced'' amplitude, i.e.~the amplitude without its dominant exponential scaling behavior, as a function of the logarithmic distance from the saturation line, i.e.~as a function of $\ln1/r^2 Q_s^2 = \rho - \rho_s$. Again we see that the difference betwen the two cases is tiny for all considered values of rapidity. Notice also that variations of the particular model were considered and still there was no change in the outcome\footnote{We mention that slightly asymmetric initial conditions, mostly resembling virtual photon - hadron scattering, were used.}.

Therefore we arrive at the conclusion that up to very high values of rapidity, the evolution with both Pomeron loop and running coupling effects included is practically the same to the one where only running coupling effects are taken into account. This is a highly nontrivial statement since, for the same initial conditions, in a fixed coupling treatment the numerical solutions show that the Pomeron loops strongly modify the results of the BK equation \cite{Soy05}.

So now it becomes natural to try to explain why the running coupling effects dominate the evolution. Let us compare the corrections induced by the Pomeron loops and the running coupling in the saturation exponent $\lambda_s$. When Pomeron loops are considered the correction is $\delta \lambda_s/\lambda_s \sim 1/R_{\rm eff}^2$, with the effective transverse space for evolution being the distance between the two boundaries; $R_{\rm eff} \sim \ln 1/\alpha$. When running coupling effects are considered the relative correction is again $\delta \lambda_s/\lambda_s \sim 1/R_{\rm eff}^2$, but now the effective transverse space for evolution is the diffusion radius; $R_{\rm eff} \sim Y^{1/6}$. So, as said earlier, it seems that loops of Pomerons might be more important. However the diffusion radius grows very slowly with rapidity\footnote{In contrast to the fixed coupling dynamics where the diffusion radius increases like $\sqrt{Y}$ and furthermore is enhanced by a big numerical coefficient.}, and what happens in practice is that there is not enough longitudinal space to become equal to (or greater than) the two-boundary width.

We might say that the final outcome is very fortunate, since the analysis of the BK equation (even in its running coupling version) which is deterministic, is much easier than the analysis of the Pomeron loop equations which represent a stochastic evolution.

\section*{Acknowledgments}

I would like to thank the organizer Christophe Royon for the invitation to lecture at the summer school. Diagrams in Figs.~\ref{FigNLO} and \ref{FigBubble} were made with JaxoDraw \cite{Jaxo}.


\end{document}